\documentclass[12pt,english]{article}
\usepackage{graphicx}
\usepackage{comment}
\usepackage{amsmath,amssymb,graphicx}
\usepackage{amsthm}
\usepackage{color}

\newtheorem{theorem}{Theorem}

\newtheorem{proposition}{Proposition}

\newcommand{\beqn}{\vspace{-0.25cm}\begin{eqnarray*}}
\newcommand{\eeqn}{\end{eqnarray*}}
\newcommand{\bneqn}{\vspace{-0.25cm}\begin{eqnarray}}
\newcommand{\eneqn}{\end{eqnarray}}

\renewcommand{\P}{\mathbb{P}}
\newcommand{\E}{\mathbb{E}}

\def\one{{\bf 1}}

\def\F{{\cal F}}

\def\P{{\mathbb P}}
\def\E{{\mathbb E}}
\def\Var{{\rm Var}\,}
\def\Cov{{\rm Cov}\,}

\def\|{\, | \,}

\def\disp{\displaystyle}

\title{Bayesian aggregation of two forecasts in the 
partial information framework}
\author{Philip Ernst, Robin Pemantle, Ville Satop{\"a}{\"a}, and Lyle Ungar}
\date{\today}

\begin{document}
\maketitle

\begin{abstract}
We generalize the results of \cite{SPU, SJPU} by showing how the Gaussian aggregator may be computed in a setting where parameter estimation is not required. We proceed to provide an explicit formula for a ``one-shot'' aggregation problem with two forecasters.
\end{abstract}

\normalsize

\section{Introduction} \label{sec:intro}

Prediction polling is a form of polling that asks a group of people to predict a common quantity. These forecasts are often used to make important decisions in medicine, economics, government, etc. In many practical settings, it is not possible to determine ex-ante which of the forecasters is the most informed or accurate (and even if this could be done, a decision to follow a specific forecaster's advice may result in relevant information from other forecasters being ignored). A more prudent solution is to pool the forecasters' information into a single consensus. This requires aggregators which can incorporate different information structures amongst the forecasters. This task motivated the work of \cite{SPU}, which introduced the Gaussian partial information framework for forecast aggregation. Further methodological framework
for estimating parameters in the Gaussian partial information model
was developed in \cite{SJPU}.  

The purpose of this letter is to further generalize the results of \cite{SPU} by showing how
the Gaussian aggregator may be computed via a (Bayesian) approach 
in which parameter estimation is not required.  Our main result
is Theorem~\ref{th:main}, which provides an explicit 
formula for the Gaussian aggregator in a ``one-shot''  (a setting in which a stream of forecasts is unavailable) aggregation 
problem with two forecasters.

In the remainder of the introduction we give a brief description 
of important challenges in event forecasting and in forecast aggregation. We proceed to
summarize the partial information framework, the Gaussian
partial information model, and our Bayesian approach.
\S \ref{sec:fixed} recalls the relevant computations
for the Gaussian model with fixed parameters.  \S \ref{sec:bayesian}
computes the Bayesian aggregator and \S \ref{sec:comparison} utilizes hypothetical data to compare the aggregators. 

\subsection{Event forecasting, loss functions, and calibration} \label{sec1}

In event forecasting, an expert is asked for a series $\{ p_n \}$
of probability forecasts for events $\{ A_n \}$.  The quantitative
study of event forecasting dates back at least three 
decades \cite{Dawi1982}, \cite{MuWi1987}. 
Usually, the expert is scored by a loss function $L(p_n , \one_{A_n})$.  
The loss function $L$ is assumed to be {\em proper}, meaning that
$p$ minimizes $\E L( \cdot ,Y)$ when $Y$ is a Bernoulli random variable
with mean $p$.  Thus a forecaster with subjective probability $p$
minimizes expected loss by forecasting $p$.  For a more complete discussion
of probability forecasting and proper loss functions, one may consult \cite{HwPe1997}.

Probability forecasts can suffer from two kinds of error: bias and imprecision.  Bias occurs when the long run frequency of 
$A_n$ for those $p_n \approx p$ is not equal to $p$.  Imprecision
occurs when $p_n$ is typically not close to zero or one.  Assuming
a sufficiently long stream of forecasts, each forecast $p_n$ may be replaced by the forecast $q(p_n)$
where $q(t)$ is the long run frequency of $A_n$ given a forecast of $t$.
The forecast is then said to be {\em calibrated}; (cf.\cite{MuWi1987})
in this work we always 
assume calibrated forecasts. Of course, there are settings in which a stream of forecasts may not be available. In such a setting it is impossible to assess bias. A reasonable protocol is to assume no bias and to encourage calibration via proper loss functions (see \cite{GJP}).

Unlike other aggregators, a distinct advantage of one-shot aggregators is their universality; they can employed when a stream of forecasts is unavailable. One-shot aggregators can also serve as an equally applicable yet a more principled alternative to common aggregators such as the average and median. The simplicity of the average and the median aggregators has long been attractive to practitioners. The key contribution of this letter is to encourage the use of more principled aggregation techniques by providing a partial information aggregator that, too, has a simple and closed-form expression.

\subsection{Forecast aggregation}
Various probability models have been implicitly or explicitly used for producing a synthesized
forecast from a collection of expert forecasts. Consider a probability space $(\Omega, \F, \P)$ and events $A \in \F$. As discussed in \cite{SPU}, an expert's forecast is considered to be calibrated if the forecast $p$ 
for an event $A$ is equal to $\P (A | \F')$ for some 
$\F' \subseteq \F$.  The $\sigma$-field $\F'$ represents the
information used to make the forecast; it need not be
the full information available to the expert.  

Some empirical work on forecast 
aggregation operates outside the above framework. For example,
the {\em measurement error framework} assumes there is a true
probability $\theta$, interpreted as the forecast made by an ``ideal''
forecaster. The actual forecasters observe a transformation
$\phi (\theta)$ together with independent mean zero idiosyncratic errors.  This leads to relatively simple aggregation rules.  For example, 
if $\phi$ is the identity, the forecasters are assumed to be 
reporting $\theta$ plus independent mean zero errors. 
The corresponding aggregator then simply averages the forecasts
\begin{equation} \label{eq:ave}
g_{\rm ave} (p_1, \ldots , p_n) := \frac{1}{n} \sum_{k=1}^n p_k \, .
\end{equation}
When the function $\phi$ is $\Phi^{-1}$ (the inverse normal CDF) this leads to 
{\em probit averaging}, defined by 
\begin{equation} \label{eq:probit}
g_{\rm probit} (p_1, \ldots , p_n) := \Phi \left (
   \frac{1}{n} \sum_{k=1}^n \Phi^{-1} (p_k) \right ) \, .
\end{equation}
Such models, while very common in practice, lead both to 
uncalibrated forecasts and suboptimal performance.  Some theoretical
problems with these models are discussed by \cite{hong-page2009};
for example, such aggregators can never leave the convex hull
of the individual expert forecasts, which is demonstrably sub-optimal
in some cases \cite{parunak2013}; see also \cite[Section~2.3.2]{SJPU}.

Letting $\F^{''}= \sigma(p_1,...,p_n)$, we define an aggregator as any random variable $\tilde{p} \in \F^{''}$. Then, amongst all such aggregators, $p^{''}$ (see (\ref{revealed}) below) is the one that minimizes the expectation of any proper loss function. It is also calibrated.

In the {\em partial information framework} for aggregation of calibrated forecasts proposed by \cite{SPU}, 
each forecaster $i$, $1 \leq i \leq N$ is assumed to have access to information $\F_i$.  
The aggregator only considers the forecasts $p_i := 
\P (A | \F_i)$.  Theoretically, the best possible forecast with this
information is the {\em revealed estimator}
\begin{equation}\label{revealed}
p^{''} := \P (A | p_i : 1 \leq i \leq N).
\end{equation}
It is clear that
$$p^{''} = g_{\rm rev} (p_1, \ldots , p_n)$$
for some function $g = g_{\rm rev}$; however, it is not possible 
to explicitly compute $g$ without making further assumptions about the model.

\subsection{Gaussian partial information model}

The Gaussian partial information model was introduced in \cite{SPU}.
The probability space $(\Omega , \F , \P)$ supports a centered
Gaussian process $\{ X_A : A \subseteq S \}$ indexed by 
the Borel subsets of a single set $S$, with $\Cov (X_A , X_B) 
= |A \cap B|$, where $| \cdot |$ refers
to Lebesgue measure.  Without loss of generality, we consider $S$
to be the unit interval. The event $A$ is defined to be the event that $X_S \geq 0$.
Each infinitesimal unit
$X_{[t,t+dt]}$ of white noise adds either positive or negative information about
the occurrence of $A$.  Each
forecaster $\F_i$ is privy to some subset of this information;
the forecaster observes all the noise in some subset $B_i \subseteq S$.  
Formally, $\F_i = \sigma (X_A : A \subseteq B_i)$.  Specification 
of the sets $\{ B_i \}$ determines the model and hence $g$. 

A number of consequences of the Gaussian partial information model are discussed
in \cite{SPU}. 
\cite[Section~5.1]{SJPU} found  that this model significantly outperformed other aggregators on the
data provided by the Good Judgment Project \cite{GJP}. The same work provides a formal mechanism for efficiently
estimating the parameters for the Gaussian partial information model. More specifically, the parameters of interest $|B_i| = \Var(X_{B_i})$ and $|B_i \cap B_j| = \Cov(X_{B_i}, X_{B_j})$ form a covariance matrix, known as the information structure, that is estimated under a specific semi-definite constraint. Note that, empirically, the exact identities of the sets $B_i$ are irrelevant; all that matters are the covariances themselves \cite{SPU}.  
For the purpose of parameter estimation, however, it is necessary
that each forecaster provides a stream of forecasts.

\indent Alternatively one may choose the parameters in advance or use a Bayesian model with a prior distribution on the unknown parameters. This letter focuses on the Bayesian approaches and generalizes the work of \cite{SJPU} by considering the scenario in which a stream of forecasts is unavailable. 

\subsection{A Bayesian approach to specifying parameters}
We now turn our focus to the problem of applying the
Gaussian partial information model in a one-shot forecasting
model. The parameters $\{ |B_i| , |B_i \cap B_j| : 1 \leq i,j \leq N \}$ 
cannot consistently be estimated because there is only one 
data point $p^{(i)}$ for each forecaster $i$.  We model this one-shot problem with a Bayesian approach;  a uniform prior $\mu$ is chosen on these 
parameters. 
Let $\nu$ denote the posterior law of the parameters given the
forecasts.  Then $p^{''}$ is the mean of $g_\alpha (p^{(1)}, 
\ldots , p^{(N)})$ when $\alpha$ is an assignment of parameters
chosen randomly from the posterior law $\nu$.

Clearly, giving an analytical solution to the problem of integrating over the space of all coherent information structures is intractable. Therefore, our subsequent discussion motivates simplifying assumptions that allow us to derive the Bayesian aggregator in closed-form. Although numerical integration can be performed trivially with Markov chain Monte Carlo (MCMC) methods, our primary motivation, as emphasized in \S \ref{sec1}, is to find a closed-form expression that encourages principled and easily available aggregation of forecasts.

\section{Aggregation function for fixed parameters} 
\label{sec:fixed}

Using the notation introduced in \S \ref{sec:intro}, we consider a model in which $N=2$, $|S| = 2$,
$|B_1| = |B_2| = 1$ and $|B_1 \cap B_2| = \rho$. Here $|S|$ can be interpreted as the total amount of information available to the forecasters. Consequently, $|B_i|$ represents the amount of information used by the $i$th forecaster. The model is invariant to scaling: one can replace $|S| = 2$ by $|S| = 2 \gamma$ and $|B_i| = 1$ by $|B_i| = \gamma$. Therefore the choice $|S| = 2$ is irrelevant \cite{SPU}. This particular choice, however, is convenient as it simplifies some of our notation. What is more important is how of much of this information each forecaster uses. As will be explained below, letting $|B_1| = |B_2| = 1$ is a non-informative and hence a natural choice. The final parameter $\rho$, treated in \cite{SPU} and in \cite{SJPU}.
as a parameter to be estimated, will later be taken to be uniform on $[0,1]$.  In this section, however, 
we fix $\rho \in [0,1]$ and compute the forecast, its marginal 
distribution, and the aggregator.

This aggregator is limited to the case of two forecasters. One way to generalize it to $N$ forecasters is to set $|B_i| = \delta$ and $|B_i \cap B_j| = \rho$ for $i = 1, \dots, N$ and $i \neq j$. This leads to a simplified Gaussian model with a compound symmetric information structure described fully by two parameters, namely $\delta$ and $\rho$. Integrating out these parameters with respect to their posterior distribution would provide a more broadly applicable aggregator. This is one of our current research projects. Unfortunately, the integrals are analytically much more challenging, and it is still not clear whether a closed-form solution exists. 

\subsection{Computing the forecast and marginals for any parameters}

A forecaster observing $X_B$ is ignorant of $X_S - X_B$ 
which is independent of $X_B$ and has distribution $N(0,|S| - |B|)$
or $\sqrt{|S| - |B|} \chi$, where $\chi$ is a standard normal.
Therefore, conditional on $X_B = x$, the forecast is 
$$p(x) 
   = \P(X_S - X_B > -x)
   = \P (\chi < (|S| - |B|)^{-1/2} x)
   = \Phi \left ( \frac{x}{\sqrt{|S| - |B|}} \right ) \, .$$

\begin{figure}[b]
\centering
\includegraphics[width=0.353\linewidth]{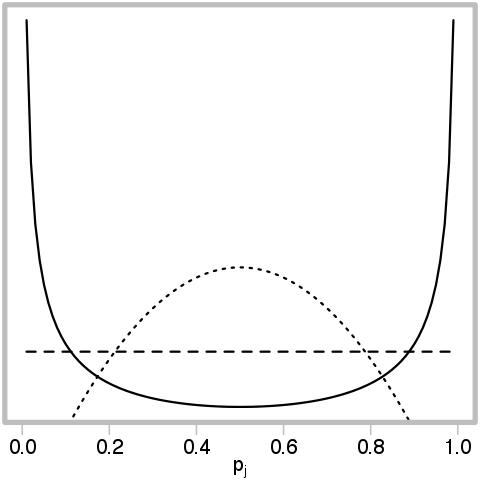}
\caption{The solid line, dashed line, and dotted line are respectively
$\beta = 7/3, 1,$ and $3/7$.
\label{fig:marginals}}
\end{figure}

Let $\beta := |B| / (|S| - |B|)$.  Because $X_B$ is distributed
as $|B|^{1/2} \chi$, we see that the law of $p$ in this model is
the law of $\Phi ( \beta^{1/2} \chi)$.  Because $\chi$ has law
$\Phi^{-1} (U)$ for $U$ uniform on $[0,1]$,
$$p(x) \sim \Phi (\beta^{1/2} \Phi^{-1} (U)) \, .$$ 
The density behaves like $(c x (1-x))^{1/\beta}$.
When $\beta < 1$ it is unimodal, when $\beta > 1$ it 
blows up at the endpoints, and when $\beta = 1$ it is 
exactly uniform (see Figure~\ref{fig:marginals}).

In this light, the choice of $|B_1| = |B_2| = |S \setminus B_1| = 
|S \setminus B_2|$ seems natural, as it causes each forecast to be
marginally uniform on $[0,1]$.

\subsection{Computing $g$ under fixed overlap}

We now specialize to the Gaussian partial 
information model $\disp |B_1| = |B_2| = \frac{|S|}{2} = 1$ and assume that the parameter $\rho = |B_1 \cap B_2|$ is known.  
We proceed to compute the aggregator. 
\begin{proposition} \label{pr:fixed}
In the Gaussian partial information model with $|B_1| = |B_2| = 1$,
$|S| = 2$ and $|B_1 \cap B_2| = \rho$, if the two experts forecast 
$p^{(1)} = p$ and $p^{(2)} = q$, then the best aggregator 
$g_\rho (p,q) := \P (A | p^{(1)} = p , p^{(2)} = q)$ is given by
\begin{equation} \label{eq:g_rho}
g_\rho (p,q) = \Phi \left ( \frac{\Phi^{-1} (p) + \Phi^{-1} (q)}
   {\sqrt{2 \rho (1 + \rho)}} \right ) \, .
\end{equation}
\end{proposition}

\noindent{\sc Proof:}
Under the Gaussian model the joint distribution of $X_S, X_{B_1}$, 
and $X_{B_2}$ is
\begin{align*}
\left(
\begin{matrix}
X_S\\
X_{B_1}\\
X_{B_2}
\end{matrix}
\right) \sim 
\mathcal{N} \left(
 \boldsymbol{0}, \left(\begin{matrix} 
\Sigma_{11} & {\bf \Sigma}_{12}\\
{\bf\Sigma}_{21} & {\bf \Sigma}_{22}\\
 \end{matrix}\right) 
\right).
\end{align*}
where
\begin{align*}
\left(\begin{matrix} 
\Sigma_{11} & {\bf \Sigma}_{12}\\
{\bf\Sigma}_{21} & {\bf \Sigma}_{22}\\
 \end{matrix}\right) 
: =
 \left(\begin{array}{c | c c }
2 & 1 & 1   \\ \hline
1 & 1 &\rho   \\ 
1 & \rho & 1  \\ 
 \end{array}\right) \, .
\end{align*}
The inverse of ${\bf \Sigma}_{22}$ is 
\begin{align*}
{\bf \Sigma}_{22}^{-1} =  \frac{1}{1-\rho^2}\left(\begin{array}{c c }
 1 &-\rho   \\ 
 -\rho & 1  \\ 
 \end{array}\right).
\end{align*}
Using the well-known properties of a conditional multivariate Gaussian 
distribution (see, e.g., \cite[Result 5.2.10]{ravishanker-dey}), 
the distribution of $X_S$ given $\boldsymbol{X} = 
(X_{B_1}, X_{B_2})'$ is 
$X_S | \boldsymbol{X} \sim \mathcal{N}(\mu_S, \sigma_S^2)$, where
\begin{align*}
\mu_S &= {\bf \Sigma}_{12} {\bf \Sigma}_{22}^{-1}\boldsymbol{X} =  
   \frac{1}{1+\rho} (X_{B_1} + X_{B_2}), \\
   \sigma_S^2 &= \Sigma_{11}-{\bf \Sigma}_{12} 
   {\bf \Sigma}_{22}^{-1}{\bf\Sigma}_{21} = \frac{2\rho}{1+\rho}.
\end{align*}
Denoting $p^{(1)}$ and $p^{(2)}$ respectively by $p$ and $q$, 
we recall that the individual forecasts are $p = \Phi(X_{B_1})$ 
and $q = \Phi(X_{B_2})$. The synthesized forecast is then
\begin{align*}
g_\rho (p,q) &= \P(X_S > 0 | p,q)\\
&= \P(X_S > 0 | X_{B_1}, X_{B_2})\\
&= 1 - \Phi\left( \frac{- \frac{1}{1+\rho} (X_{B_1} + X_{B_2})}{\sqrt{\frac{2\rho}{1+\rho}}} \right)\\
&=\Phi\left( \frac{\Phi^{-1}(p) + \Phi^{-1}(q)}{\sqrt{2\rho(1+\rho)}} \right).
\end{align*}
$\hfill \Box$

\section{Bayesian model} \label{sec:bayesian}

We now further develop our model in \S \ref{sec:fixed} via a Bayesian approach.
We assume that the overlap parameter $\rho$ has a
prior distribution that is uniform over the interval $[0,1]$.  The posterior
distribution is not uniform because the likelihood
$$\lambda_\rho (p,q) := \P(p,q \| \rho)$$
of $(p,q)$ given $\rho$ is nonconstant, whence Bayes' Rule
applied with the uniform prior gives a nonconstant posterior.
Given $p$ and $q$, posterior probabilities are given by quotients
of integrals:
\begin{eqnarray} \label{eq:P}
g(p,q) := \P (A \| p , q) & = & \int \P (A \| p , q , \rho) \,  \, 
   \P (\rho \| p , q) \\[2ex]
& = & \frac{\int f(p,q,\rho) \lambda_\rho (p,q) \, d\rho}
   {\int \lambda_\rho (p,q) \, d\rho} \, .  \nonumber
\end{eqnarray}
Here we include a factor of $\int \lambda_\rho \, d\rho$ in the 
denominator so that we may, if we choose, allow $\lambda_\rho$ 
not to be normalized to have total mass one. 

\begin{theorem} \label{th:main}
$g(p,q)$ can be evaluated in the following closed-form expression
\begin{equation} \label{eq:simple}
g(p,q) = \left \{ \begin{array}{ll}
   \displaystyle{\frac{p - (1 - 2q)}{2q}} & \;\; p > \max \{ q , 1-q \} \\
   \displaystyle{\frac{p}{2(1-q)}} & \;\; p < \min \{ q , 1-q \} \\[2ex]
   \displaystyle{\frac{q - (1 - 2p)}{2p}} & \;\; q > \max \{ p , 1-p \} \\
   \displaystyle{\frac{q}{2(1-p)}} & \;\; q < \min \{ p , 1-p \}.
\end{array} \right.
\end{equation}
\end{theorem}

\noindent{\sc Proof:}
To compute $\lambda_\rho (p,q)$, recall that $Z_1$ and $Z_2$ are
standard normals with covariance $\rho$ and that $(Z_1 , Z_2)$
maps to $(p,q)$ by $\Phi$ in each coordinate.  The density of
$(Z_1 , Z_2)$ at $(x,y)$ is proportional to   
	$$(2 \pi)^{-1} (\det Q)^{1/2} \exp \left [ \frac{1}{2} Q(x,y) \right ],$$
where the quadratic form $Q$ is the inverse of the covariance
matrix
$$Q = \frac{1}{1 - \rho^2} \left [ \begin{array}{cc} 
   1 & - \rho \\ - \rho & 1 \end{array} \right ] \, .$$
Thus the density $h(x,y)$ of $(Z_1 , Z_2)$ at $(x,y)$ is equal to
\begin{equation} \label{eq:xy}
\frac{1}{2 \pi} (1 - \rho^2)^{-1/2} \exp \left [ -
   \frac{x^2 + y^2 - 2 \rho xy}{2 (1 - \rho^2)} \right ] \, .
\end{equation}
The Jacobian of the map $(x,y) \mapsto (\Phi (x) , \Phi (y))$ 
at $(x,y)$ is given by 
\begin{equation} \label{eq:J}
\frac{1}{2 \pi} \exp \left [ - \frac{1}{2} \left ( x^2 + y^2 \right ) 
   \right ]
\end{equation}
and therefore 
\begin{eqnarray*}
\lambda_\rho (p,q) & = & \left. h (x,y) J(x,y)^{-1} 
   \right |_{x = \Phi^{-1} (p) , y = \Phi^{-1} (q)} \\[2ex]
& = & c (1 - \rho^2)^{-1/2} \exp \left [ - \frac{\rho^2 x^2 
   - 2 \rho x y + \rho^2 y^2}{2 (1 - \rho^2)} \right ] \, .
\end{eqnarray*}
Combining this with ~\eqref{eq:g_rho} and ~\eqref{eq:P} gives
\begin{equation} \label{eq:post}
g(p,q) = \frac{\displaystyle{\int \Phi \left ( \frac{\Phi^{-1} (p) + \Phi^{-1} (q)}
   {\sqrt{2 \rho (1 + \rho)}} \right ) 
   (1 - \rho^2)^{-1/2} \exp \left [ - \frac{\rho^2 \Phi^{-1} (p)^2
   - 2 \rho \Phi^{-1} (p) \Phi^{-1} (q) 
   + \rho^2 \Phi^{-1} (q)^2}{2 (1 - \rho^2)} \right ] \, d\rho}}
   {\displaystyle{(1 - \rho^2)^{-1/2} \exp \left [ 
   - \frac{\rho^2 \Phi^{-1} (p)^2 - 2 \rho \Phi^{-1} (p) \Phi^{-1} (q) 
   + \rho^2 \Phi^{-1} (q)^2}{2 (1 - \rho^2)} \right ]}} \, .
\end{equation} 

\noindent By symmetry, we may assume without loss of generality that $p < q$.
Removing a factor of 
\begin{equation}
\disp \exp \left [ \frac{1}{2} \left (
\Phi^{-1} (p)^2 + \Phi^{-1} (q)^2 \right ) \right ]
\end{equation}
from both the numerator and denominator of~\eqref{eq:post} gives
\begin{equation}
g(p,q) = \frac{\disp {\int_{0}^{1}\Phi\left(\frac{\Phi^{-1}(p)
   + \Phi^{-1}(q)}{\sqrt{2\rho(1+\rho)}}\right)
     \frac{1}{\sqrt{1-\rho^2}} 
     \exp \left ( -\frac{\Phi^{-1}(p)^2-2\rho \Phi^{-1}(p)\Phi^{-1}(q)
     + \Phi^{-1}(q)^2}{2(1-\rho^2)} \right ) \, d\rho}}
     {\disp {\int_{0}^{1} \hspace{1.55in} \frac{1}{\sqrt{1-\rho^2}} 
     \exp \left ( -\frac{\Phi^{-1}(p)^2-2\rho \Phi^{-1}(p)\Phi^{-1}(q)
    + \Phi^{-1}(q)^2}{2(1-\rho^2)} \right ) d\rho}} \, . \label{eq:g}
\end{equation}
We first compute the denominator of \eqref{eq:g} and then proceed to compute the numerator.

\subsection{Computation of the denominator}

Let us denote the denominator of~\eqref{eq:g} by 
\begin{equation} \label{eq:I_2}
I_2 := \int_{0}^{1}\frac{1}{\sqrt{1-\rho^2}} 
   \exp \left ( -\frac{\Phi^{-1}(p)^2-2\rho \Phi^{-1}(p)\Phi^{-1}(q)
   + \Phi^{-1}(q)^2}{2(1-\rho^2)} \right ) \, d\rho \, .
\end{equation}
Let the density, CDF and tail of the bivariate standard normal 
with correlation parameter $\rho \in (-1, 1)$ be defined respectively by 
\begin{eqnarray*}
\phi_2(x,y;\rho) & = & \frac{1}{2\pi\sqrt{1-\rho^2}}
   e^{-\frac{x^2-2\rho xy+y^2}{2(1-\rho^2)}} \\[2ex]
\Phi_2(b_1,b_2;\rho) & = & 
   \int_{-\infty}^{b_1}\int_{-\infty}^{b_2}\phi_2(x,y;\rho)dydx \\[2ex]
L(b_1 , b_2 , \rho) & = & \Phi_2 (-b_1, -b_2, \rho) \, .
\end{eqnarray*}
Plackett's formula (\cite{plackett1954}) gives that 
$$\frac{\partial L(b_1, b_2, \rho)}{\partial \rho} = 
   \frac{\exp \left ( \disp{ -\frac{b_1^2-2\rho b_1b_2+b_2^2}{2(1-\rho^2)}}
    \right ) } 
   {2\pi\sqrt{1-\rho^2}}$$
specializes to the integrand in~\eqref{eq:I_2} when $b_1 = \Phi^{-1} (p)$
and $b_2 = \Phi^{-1} (q)$, whence
$$I_2 = \int_0^1 2 \pi \frac{\partial}{\partial \rho}
   L \left ( \Phi^{-1} (p) , \Phi^{-1} (q) , \rho \right )
   \, d\rho \, .$$
Assuming $p < q$ and utilizing the identities $L(b_1, b_2 , 0) = \Phi (-b_1) \Phi (-b_2)$ and 
$L(b_1, b_2 , 1) = \Phi (-\max \{ b_1 , b_2 \})$, we obtain
\begin{eqnarray}
I_2 & = & 
2 \pi \left [ L (\Phi^{-1}(p), \Phi^{-1}(q), 1) 
   - L (\Phi^{-1}(p), \Phi^{-1}(q), 0) \right ] \nonumber \\[1ex]
& = & 2 \pi \left [ \Phi (- \max \{ \Phi^{-1} (p) , \Phi^{-1} (q) \} ) 
   - \Phi ( - \Phi^{-1} (p)) \Phi (- \Phi^{-1} (q)) \right ] \nonumber \\[1ex]
& = & 2 \pi (1-q) p \, . \label{eq:I_2final}
\end{eqnarray}

\subsection{Computation of the numerator}

\begin{figure}[b]
\centering
\includegraphics[width=2.0in]{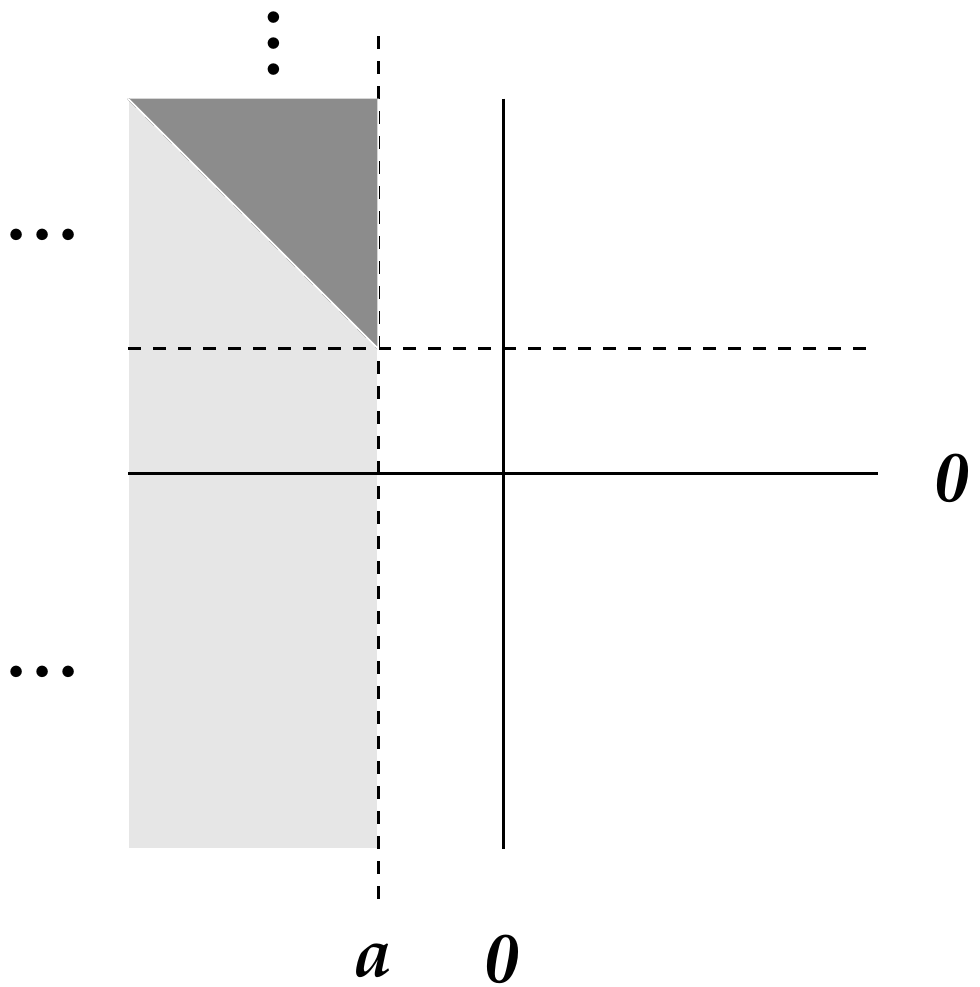}
\caption{The darker region has probability $\P(Y_1 \leq a)^2 / 2$.}
\label{fig:meyer}
\end{figure}

We denote the numerator of ~\eqref{eq:g} as
\begin{equation} \label{eq:I_1}
I_1 := \int_{0}^{1}
   \Phi \left ( \frac{\Phi^{-1}(p)+\Phi^{-1}(q)}{\sqrt{2\rho(1+\rho)}} \right )
   \frac{1}{\sqrt{1-\rho^2}} 
   \exp \left ( 
   - \frac{\Phi^{-1}(p)^2-2\rho \Phi^{-1}(p)\Phi^{-1}(q)+\Phi^{-1}(q)^2}
   {2(1-\rho^2)} \right ) d\rho \, .
\end{equation} 
Extending previous notation, we denote the trivariate normal CDF by
\begin{equation}\label{eqn:tncdf}
\Phi_3(b_1,b_2, b_3; R) = 
   \frac{1}{(2\pi)^{3/2}|R|^{1/2}}
   \int_{-\infty}^{b_1}\int_{-\infty}^{b_2}\int_{-\infty}^{b_3}
   \exp \left ( -\frac{x^{T} R^{-1} x}{2} \right ) \, dx_3\, dx_2\, dx_1,
\end{equation}
where $R=(\rho_{ij})$ is the correlation matrix.  \cite{plackett1954} 
contains a formula for the partial derivative of 
the trivariate CDF with respect to the coefficient $\rho_{12}$,
meaning that the $(1,2)$ and $(2,1)$ entries of $R$ change while 
all other entries remain constant:
\begin{equation}\label{eqn:derivfi3}
\frac{\partial \Phi_3(b_1, b_2, b_3;R)}{\partial \rho_{12}}
   = \frac{\exp \left ( -\frac{b_1^2-2\rho_{12} b_1b_2+b_2^2}{2(1-\rho^2)}   
     \right ) }{2\pi\sqrt{1-\rho_{12}^2}}
   \Phi(u_3(\rho_{12})) \, ,
\end{equation}
where
\begin{equation} \label{eq:u}
u_3(\rho) =
   \frac{b_3(1-\rho^2)-b_1(\rho_{31}-\rho\rho_{32})
   - b_2(\rho_{32}-\rho\rho_{31})}
   {\sqrt{(1-\rho^2)(1-\rho^2-\rho_{31}^2-\rho_{32}^2
   + 2\rho\rho_{31}\rho_{32})}} \, .
\end{equation}
Plugging in 
$$
b_1 = -\Phi^{-1}(p), \;\; b_2 = -\Phi^{-1}(q), \;\; b_3 = 0, \;\; 
   \mbox{ and } \rho_{31} = \rho_{32} = \frac{1}{\sqrt{2}}
$$
gives
$$
u_3 (\rho_{12}) = 
   \frac{\disp \frac{1-\rho_{12}}{\sqrt{2}} 
      \left (\Phi^{-1} (p) + \Phi^{-1} (q) \right )}
   {\sqrt{(1-\rho_{12}^2) (\rho_{12} - \rho_{12}^2)}}
   = \frac{\Phi^{-1} (p) + \Phi^{-1} (q)}{\sqrt{2 \rho_{12} (1 + \rho_{12})}}.
$$
This leads to
\begin{eqnarray} 
\frac{\disp \partial \Phi_3 \left ( -\Phi^{-1} (p) , -\Phi^{-1} (q) , 0 ;
   \left ( {\tiny \begin{array}{ccc} 
   1 & \rho_{12} & \sqrt{1/2} \\ 
   \rho_{12} & 1 & \sqrt{1/2} \\ 
   \sqrt{1/2} & \sqrt{1/2} & 1 \end{array}} \right ) \right ) }
   {\disp \partial \rho_{12}} && \nonumber \\[3ex]
= \;\; \frac{\exp \left ( 
    - {\disp \frac{\Phi^{-1} (p)^2 - 2 \rho_{12} \Phi^{-1} (p) \Phi^{-1} (q) 
          + \Phi^{-1} (q)^2}
           {2(1-\rho_{12}^2)}   }
     \right ) }{2\pi\sqrt{1-\rho_{12}^2}}
   \; \Phi \left ( 
   \frac{\Phi^{-1} (p) + \Phi^{-1} (q)}{\sqrt{2 \rho_{12} (1 + \rho_{12})}} 
    \right )  \, . \label{eq:plugged} &&
\end{eqnarray}
Integrating~\eqref{eq:plugged} as $\rho_{12}$ ranges from~0 to~1 and
comparing with \eqref{eq:I_1}, we see that
\begin{eqnarray} 
I_1 & = & 2 \pi \int_0^1 \frac{\partial}{\partial \rho_{12}} 
   \Phi_3 \left ( -\Phi^{-1} (p) , -\Phi^{-1} (q) , 0 ;
   \left ( {\tiny \begin{array}{ccc} 
   1 & \rho_{12} & \sqrt{1/2} \\ 
   \rho_{12} & 1 & \sqrt{1/2} \\ 
   \sqrt{1/2} & \sqrt{1/2} & 1 \end{array}} \right ) \right )
   \, d\rho_{12} \nonumber \\[2ex]
& = & 2 \pi \left [ 
     \Phi_3 \left ( - \Phi^{-1}(p), -\Phi^{-1}(q), 0 ; R \right)
   - \Phi_3 \left ( - \Phi^{-1}(p), -\Phi^{-1}(q), 0;R^* \right ) \right ] 
   \, , \label{eq:int1}
\end{eqnarray}
where the matrices $R, \ R^*$ are given by
\begin{equation}\label{eqn:RRast}
R=\left( \begin{array}{ccc}
             1 & 1 & \frac{1}{\sqrt{2}} \\
             1 & 1 &\frac{1}{\sqrt{2}} \\
             \frac{1}{\sqrt{2}} & \frac{1}{\sqrt{2}} & 1 \\
           \end{array}
         \right),
         \qquad R^{\ast}=\left(
           \begin{array}{ccc}
             1 & 0 & \frac{1}{\sqrt{2}} \\
             0& 1 & \frac{1}{\sqrt{2}} \\
             \frac{1}{\sqrt{2}} & \frac{1}{\sqrt{2}} & 1 \\
           \end{array}
         \right).
\end{equation}
Computing $\Phi_3(-\Phi^{-1}(p), -\Phi^{-1}(q), 0; R)$,
we note that $R$ forces $X_1 = X_2$, whence 
\begin{equation}
\Phi_3 (a,b,c;R) = \Phi_2 (-\max \{ a,b \} , c ; R')
\end{equation}
where $R' = \left ( \begin{array}{cc} 1 & \sqrt{1/2} \\ \sqrt{1/2} & 1 
\end{array} \right )$.  If $(X_1, X_2)$ is Gaussian with covariance $R'$
then $X_1 = Y_1$ and $X_2 = (Y_1 + Y_2) / \sqrt{2}$ where $(Y_1, Y_2)$
are independent standard normals.  Thus, if $p < q$, 
\begin{eqnarray*} 
\Phi_3 \left ( - \Phi^{-1}(p), -\Phi^{-1}(q), 0 ; R \right)
   & = & \Phi_2 \left ( - \Phi^{-1} (q) , 0 ; R' \right ) \\
   & = & \P (X_1 \leq - \Phi^{-1} (q) , X_2 \leq 0) \\
   & = & \P (Y_1 \leq - \Phi^{-1} (q) , Y_2 \leq - Y_1) \, .
\end{eqnarray*}
\cite{meyer2009} remarks (see Figure \ref{fig:meyer}) that 
$$\P (Y_1 \leq a, Y_2 \leq - Y_1) = \P (Y_1 \leq a) - \frac{1}{2}
   \P (Y_1 \leq a)^2 \, .$$

\noindent Thus,
\begin{equation} \label{eq:R}
\Phi_3 \left ( - \Phi^{-1}(p), -\Phi^{-1}(q), 0 ; R \right)
   = (1-q) - \frac{(1-q)^2}{2} \, .
\end{equation}

\begin{figure}[b]
\centering
\includegraphics[width=3in]{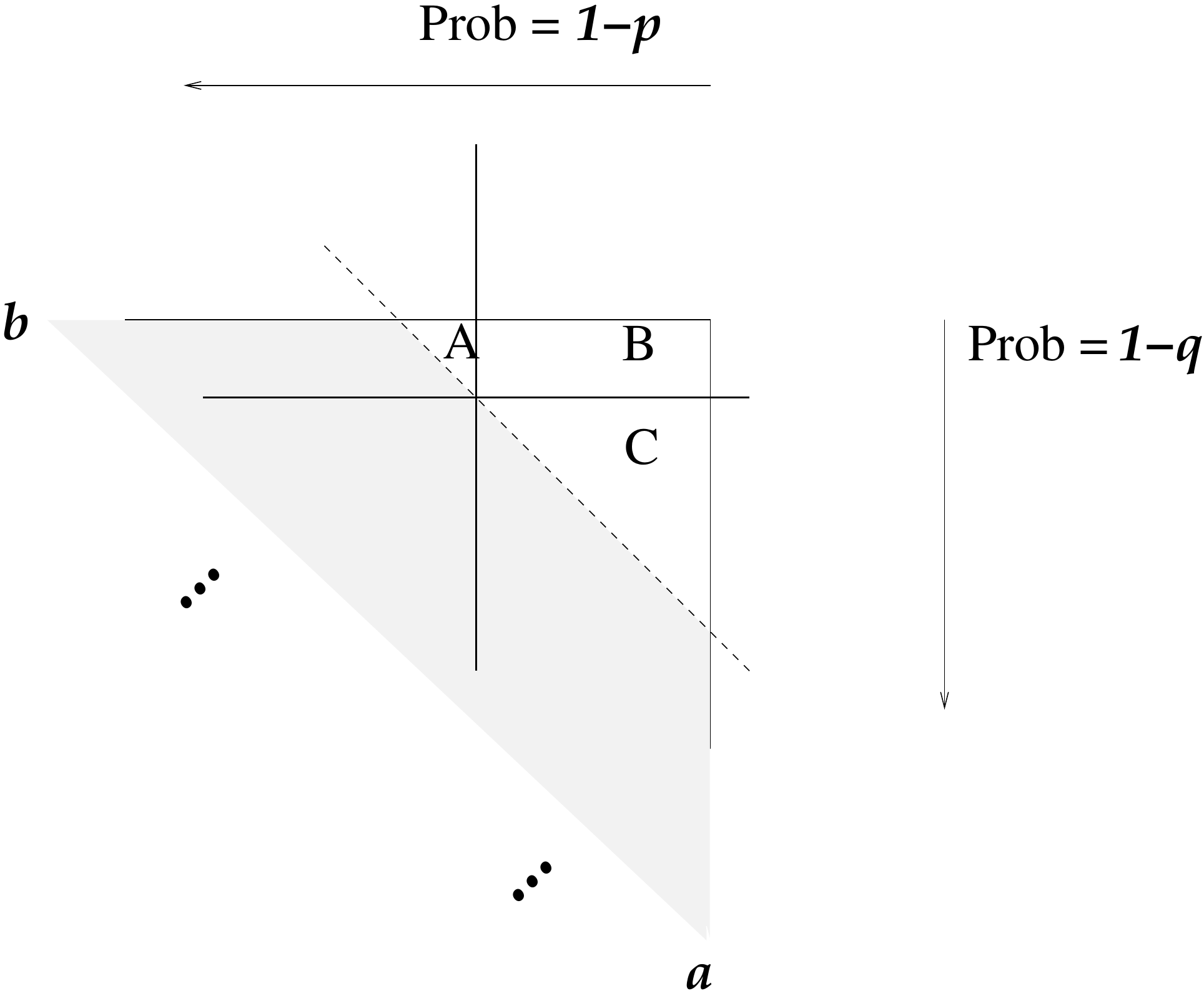}
\caption{The area of quadrant $\{ Y_1 \leq a, Y_2 \leq b \}$ is $(1-p)(1-q)$.  
We subtract from this areas $A, B$ and $C$, which are respectively
$(1/2 - p)^2/2, (1/2 - p)(1/2 - q)$, and $(1/2 - q)^2 / 2$.}
\vspace{1in}
\label{fig:dissection}
\end{figure}

We next compute $\Phi_3\left(-\Phi^{-1}(p), -\Phi^{-1}(q), 0;R^* \right)$.
In this case, $(X_1, X_2, X_3) = (Y_1, Y_2, (Y_1 + Y_2)/\sqrt{2})$ where
again $(Y_1, Y_2)$ is a pair if independent standard normals.  We
therefore must compute
$$\P (Y_1 \leq - \Phi^{-1} (p), Y_2 \leq -\Phi^{-1} (q), Y_1 + Y_2 \leq 0).$$  
We claim that
\begin{equation} \label{eq:cases}
\P (Y_1 \leq - \Phi^{-1} (p), Y_2 \leq -\Phi^{-1} (q), Y_1 + Y_2 \leq 0)
   = \begin{cases} (1-p)(1-q) & \mbox{if } p+q \geq 1; \\
   {\disp \frac{1-p^2-q^2}{2}} & \mbox{if } p+q < 1 .\end{cases} 
\end{equation}
When $p+q \geq 1$, then $Y_1 \leq - \Phi^{-1} (p)$ and 
$Y_2 \leq -\Phi^{-1} (q)$ together imply $Y_1 + Y_2 \leq 0$.
Thus the probability is $\Phi_2 (- \Phi^{-1} (p) , - \Phi^{-1} (q))
- (1-p)(1-q)$ as claimed.
When $p+q < 1$, the claimed result follows 
as illustrated in Figure~\ref{fig:dissection}.

\noindent Plugging in \eqref{eq:R} and~\eqref{eq:cases} into \eqref{eq:int1} yields
$$I_1 = \begin{cases}
   2 \pi \left [ (1-q) - \frac{(1-q)^2}{2} - (1-q)(1-p) \right ]       
      & \mbox{if } p < q \mbox{ and } p+q \geq 1 \\
   2 \pi \left [ (1-q) - \frac{(1-q)^2}{2} - \frac{1-p^2-q^2}{2} \right ] 
      & \mbox{if } p < q \mbox{ and } p + q \leq 1.
        \end{cases} $$
Finally, dividing by $I_2$ gives our desired result:
$$g(p,q) = \begin{cases}
   {\disp \frac{(1-q)-\frac{(1-q)^2}{2}-(1-q)(1-p)}{p(1-q)}=\frac{q-(1-2p)}{2p}}
      & \mbox{if } p < q \mbox{ and } p+q \geq 1 \\[2ex]
   {\disp \frac{(1-q)-\frac{(1-q)^2}{2}-\frac{1-p^2-q^2}{2}}{p(1-q)}=\frac{p}{2(1-q)}}
      & \mbox{if } p < q \mbox{ and } p + q \leq 1 . \end{cases}$$

\section{Comparison of aggregations with hypothetical data} \label{sec:comparison}

We now offer a concrete comparison amongst aggregators. Let us suppose two experts forecast respective 
probabilities $p_1 = 0.6$ and $p_2 = 0.8$.  We wish to consider a number of
of aggregators.  The first two were discussed in~\eqref{eq:ave} and 
\eqref{eq:probit}, namely the simple average 
$p^{\rm ave} := g_{\rm ave} (p_1,p_2)$ and the inverse-phi average
$p^{\rm probit} := g_{\rm probit} (p_1,p_2)$.  As discussed previously,
these values are constrained to lie between $p_1$ and $p_2$.  

We compare the revealed forecast to these two aggregators and to two aggregators others not 
constrained to the convex hull.  The first of the two latter aggregators is from
Gaussian model with fixed overlap parameter $\rho = 1/2$.  
The second is the log odds summing aggregator.  The log odds summing aggregator,
which we have not discussed above, is based on the 
probability model in which each forecaster begins with a prior probability 
estimate of $p = 1/2$ (equivalently $\log (p/(1-p)) = 0$ and observes the result of an independent experiment.

By Bayes rule, this experiment affects the posterior probability
by an additive increment in the log odds.  The result of the two
independent experiments is to add both increments to the log odds,
resulting in an estimator $p^{\rm log \; odds}$ which is the most 
extreme of those we have considered.  Just as $p^{\rm ave}$ and 
$p^{\rm probit}$ are demonstrably underconfident, $p^{\rm log \; odds}$
is overconfident because it assumes that the experts' data are
completely disjoint.
Below we present the following values for the various synthesized forecasts 
(rounded to the nearest $0.001$).  

\begin{center}
\begin{tabular}{|c|c|}
\hline
$p^{\rm ave}$ & $0.700$  \\
\hline
$p^{\rm probit}$ & $0.708$ \\
\hline
$p^{1/2}$ & $0.814$ \\
\hline
$p^{\rm revealed}$ & $0.833$ \\
\hline
$p^{\rm log \; odds}$ & $0.857$\\
\hline
\end{tabular}
\end{center}

The range of values
of these aggregators is quite broad, extending from
$7/10$ at the low end to $6/7$ at the high end. \footnote{A number 
of these aggregators give
rational values on rational inputs.} Almost anyone
in the business, if given forecasts of $3/5$ and $4/5$,
would place their estimate between $7/10$ and $6/7$.
The choice of model substantially alters the particular
aggregate forecast within the interval of plausible forecasts,
and is therefore quite important.  We also remark that this
choice is not a mathematical one but a practical one.  Different
forecasting problems may call for different aggregation techniques.

\begin{figure}[h]
\centering
\includegraphics[scale=.40]{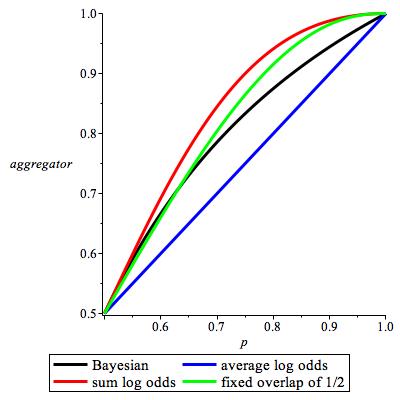} \;\;\;\;\;\;\;\;\;\;
\includegraphics[scale=.40]{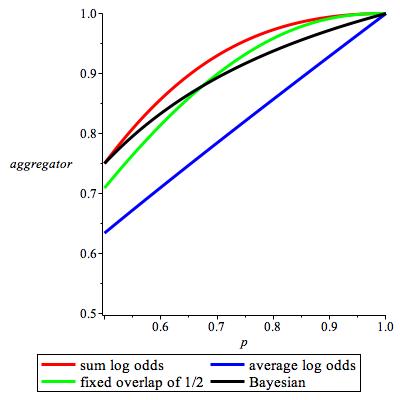}
\caption{Graphical comparisons of aggregators.}
\label{fig:plots}
\end{figure}

The left graph in Figure~\ref{fig:plots} provides a visual comparison 
of the above synthesis functions by graphing the diagonal values,
that is those where $p = q$.  By symmetry, it suffices
to graph each of these on the interval $[1/2,1]$.  When
$p = q = x$, both the average $p^{\rm ave}$ and the inverse-phi average
$p^{\rm probit}$ are also equal to $x$; these are shown by 
the blue line.  The red curve is $p^{\rm log odds}$,
which is always greatest of the aggregators under consideration.
The green and black curves represent
$p^{1/2}$ and $p^{\rm revealed}$ respectively, which are the
two partial overlap models.  As is evident, these are not
strictly ordered.  On the right, graphs are shown for
$p \in [1/2,1]$ and $q = (1+p)/2$.  When $p \neq q$, as in
the figure on the right, the inverse-phi average (shown in brown)
is distinct from the average.

One final remark concerns $p^{\rm probit}$, a popular choice 
for empirically driven aggregators.  While it may seem atheoretical,
in fact it arises as the limit as $\rho \to 1$ of the fixed
overlap aggregator.  To see this, denote the values of $X_S$ as 
$S$ varies over the algebra of sets generated by $B_1$ and $B_2$ by
$U := X_{B_1 \setminus B_2}$, $V := X_{B_2 \setminus B_1}$,
$M := X_{B_2 \cap B_1}$ and $W := X_{(B_2 \cup B_1)^c}$; thus
$X_{B_1} = U + M, X_{B_2} = V + M$ and $X_S = U+V+M+W$,
where $U,V,M,W$ are independent Gaussians with
respective variances $1-\rho, 1-\rho, \rho, \rho$.

\begin{figure}[h!]
\centering
\includegraphics[scale=.40]{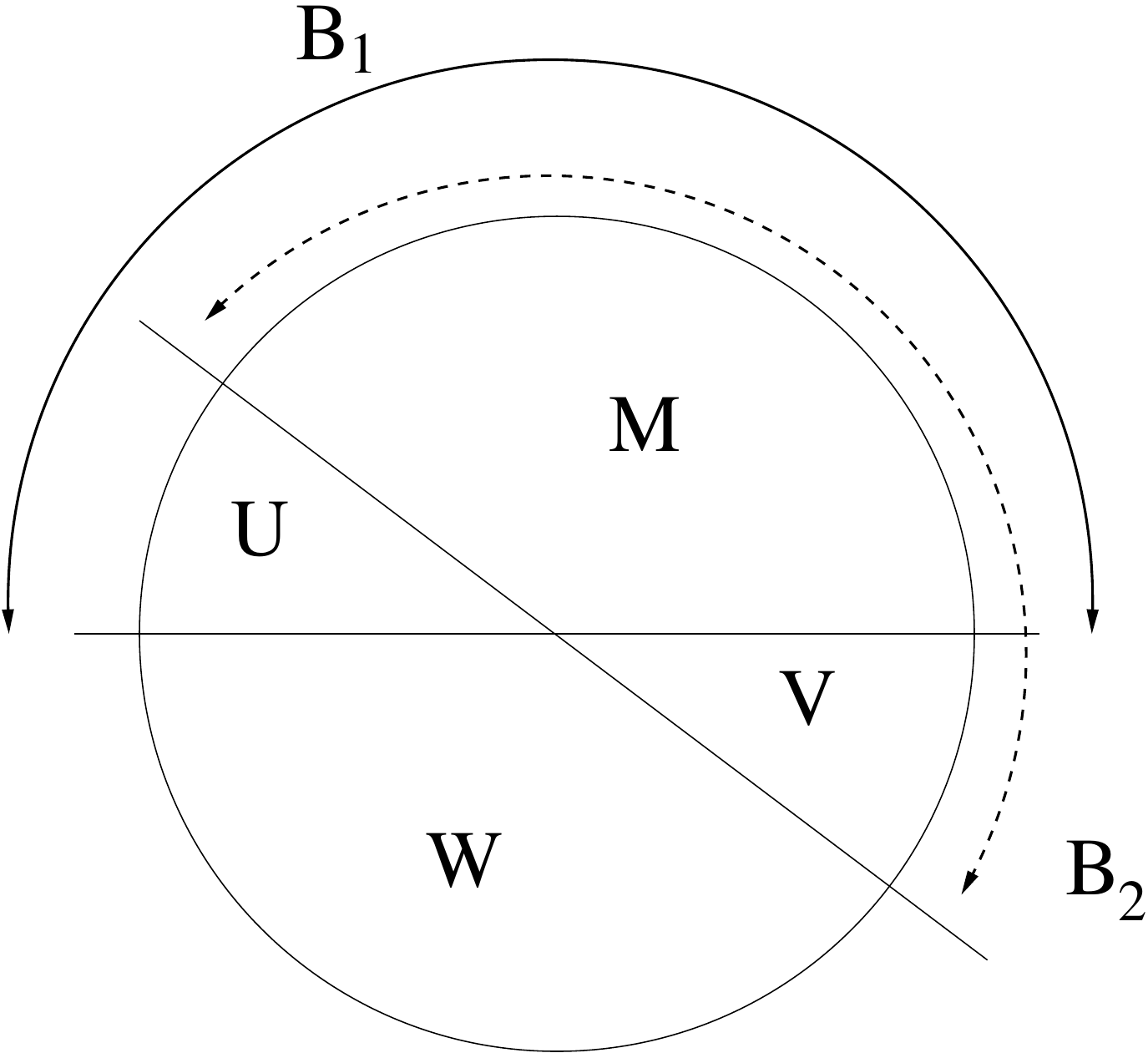}
\caption{Information partition: forecaster $i$ sees white noise in the
region $B_i$, with normalized area $1/2$; the overlap of information sets
is $M$; the symmetric difference is $U \cup V$.}
\label{fig:overlap}
\end{figure}

As $|U| = |V| \to 0$ in Figure~\ref{fig:overlap}, asymptotically, the 
likeliest way to achieve $U+M = a$ and $V+M = b$ is to let 
$M = (a+b)/2$ and $U = -V = (a-b)/2$.  These choices become 
forced in the limit.  
Applying this with $a = \Phi^{-1} (p)$ and $b = \Phi^{-1} (q)$ 
shows that the revealed forecast is $\Phi ((a+b)/2)$ which is
the inverse-phi average.  In other words, this forecast is practical only if we have reason to believe that both forecasters know nearly all information possible and that they find highly 
relevant information in the small part of their information 
that is not shared.

\newpage

\noindent \textbf{Acknowledgments}\\
This research was supported in part by NSF grant \# DMS-1209117 
and a research contract to the University of Pennsylvania and the 
University of California from the 
Intelligence Advanced Research Projects Activity (IARPA) 
via the Department of Interior National Business Center 
contract number D11PC20061.  The U.S. Government is authorized 
to reproduce and distribute reprints for Government purposes 
notwithstanding any copyright annotation thereon. 
Disclaimer: The views and conclusions expressed herein are
those of the authors and should not be interpreted as necessarily
representing the official policies or endorsements, either expressed
or implied, of IARPA, DoI/NBC, or the U.S. Government.

\newpage

\bibliography{mybibfile}

\begin{thebibliography}{99}

\bibitem[Dawid (1982)]{Dawi1982}
Dawid, A.
\emph{The well-calibrated Bayesian.},
Journal of the American Statistical Association.  \textbf{77} (1982), 605--610.

\bibitem[Hong and Page (2009)]{hong-page2009}
Hong, L. and Page, S.
\emph{Interpreted and generated signals.},
Journal of Economic Theory.  \textbf{144} (2009), 2174--2196.


\bibitem[Hwang and Pemantle (1997)]{HwPe1997}
Hwang, J.T. and Pemantle, R.
\emph{Estimating the truth of an indicator function of a statistical
hypothesis under a class of proper loss functions.},
Statistics \& Decisions.  \textbf{15} (1997), 103--128.

\bibitem[Meyer (2009)]{meyer2009}
Meyer, C.
\emph{The bivariate normal copula.}
Communications in Statistics--Theory and Methods.  \textbf{42} (2009), 2402--2422.


\bibitem[Murphy and Winkler (1987)]{MuWi1987}
Murphy, A. and Winkler, R.
\emph{ A general framework for forecast verification.},
Monthly Weather Review.  \textbf{115} (1987), 1330--1338.

\bibitem[Parunak et al. (2013)]{parunak2013}
Parunak, H., Brueckner, S., Hong, L., Page, S. and Rohwer, R.
\emph{Characterizing and aggregating agent estimates},
In Proceedings of the 2013 International Conference on Autonomous Agents and Multi-agent Systems, pages 1021--1028. International Foundation for Autonomous Agents and Multi-agent Systems, Richland, SC. (2013). 

\bibitem[Plackett (1954)]{plackett1954}
Plackett, R.
\emph{A reduction formula for normal multivariate integrals.},
Biometrika.  \textbf{41} (1954), 351--360.


\bibitem[Ravishanker and Dey (2001)]{ravishanker-dey}
Ravishanker, N. and Dey, D.
\emph{A First Course in Linear Model Theory.}
CRC Press (2001).


\bibitem[Satop{\"a}{\"a} et al. (2014)]{SPU}
Satop{\"a}{\"a}, V. and Pemantle, R. and Ungar, L.
\emph{Modeling probability forecasts via information diversity,},
To appear in Journal of the American Statistical Association. (2015).

\bibitem[Satop{\"a}{\"a} et al. (2015)]{SJPU}
Satop{\"a}{\"a}, V., Jensen, S., Pemantle, R., and Ungar, L.
\emph{Partial information framework: aggregating estimates from diverse information sources.},
\textit{Preprint}. (2015).

\bibitem[Ungar et al. (2012)]{GJP}
Ungar, L., Mellers,B., Satop{\"a}{\"a}, V., Tetlock, P. and Baron, J.
\emph{The Good Judgment Project: A Large Scale Test of 
      Different Methods of Combining Expert Predictions},
In The Association for the Advancement of Artificial Intelligence 2012 Fall Symposium Series. (2012).


\end{thebibliography}

\end{document}